\documentclass[preprint,aip,showpacs,preprintnumbers,amsmath,amssymb]{revtex4-1}
\usepackage{graphicx}
\usepackage{dcolumn}
\usepackage{bm}
\usepackage{float}

\draft

\begin{document}

\title{Local transport measurements on epitaxial graphene}

\author{J. Baringhaus} \author{F. Edler}
\affiliation{Institut f\"ur Festk\"orperphysik, Leibniz Universit\"at
Hannover, Appelstra\ss e 2, 30167 Hannover, Germany}

\author {C. Neumann} \author{C. Stampfer}
\affiliation{JARA-FIT and II. Institute of Physics A, RWTH Aachen University, 52074 Aachen, Germany}

\author{S. Forti} \author{U. Starke}
\affiliation{Max-Planck-Institut f\"ur Festk\"orperforschung, Heisenbergstrasse 1, 70569 Stuttgart, Germany}

\author{C. Tegenkamp} \email{tegenkamp@fkp.uni-hannover.de}
\affiliation{Institut f\"ur Festk\"orperphysik, Leibniz Universit\"at Hannover, Appelstra\ss e 2, 30167 Hannover, Germany}

\date{\today}

\begin{abstract}
Growth of large-scale graphene is still accompanied by imperfections.
By means of a four-tip STM/SEM the local structure of graphene grown on SiC(0001) was correlated with scanning electron microscope images and spatially resolved transport measurements. The systematic variation of probe spacings and substrate temperature has clearly revealed two-dimensional transport regimes of Anderson localization as well  as of diffusive transport. The detailed analysis of the temperature dependent data demonstrates that the local on-top nano-sized contacts do not induce significant strain to the epitaxial graphene films.
\end{abstract}
\maketitle

Graphene has been intensively studied over the last  years. Much effort has been spent in optimizing the growth towards almost defect-free large-scale graphene, which is important if graphene is intended to be used in device structures or electronic circuits \cite{Novoselov12,Emtsev09, Reina09}. Nonetheless, in all growth processes used so far residual imperfections are present which even may limit the transport properties of the intact graphene areas.

Transport properties in low dimensional structures are often probed in a  large-scale geometry using macroscopic contacts. Usually, these contacts  are fabricated by means of complicated lithographic processes,  thus easily  parasitic effects can be induced which may modify the properties of the actual structure \cite{Berdebes11}. Furthermore, transport is a non-local technique, therefore, this approach lacks the control of relevant imperfections which for instance can  give rise to potential fluctuations across the sample \cite{Martin08,Miller09}.

Against this background nano-scale transport with defined contacts and full control of the morphology is tempting. Keeping in mind that the chemical potentials at the contact sites are not only tuned by  external fields but also due to interface effects the ultimate control of a  transport measurement is synonymous with the control of the contacts and interfaces with respect to their structure, size, work function, etc. This idea has been already  picked up successfully for scanning probe techniques  by using functionalized tips for high resolution experiments on molecules. \cite{Weiss10,Gross11, Welker12}.

In this Letter we concentrate on spatially resolved transport properties of monolayer graphene grown on 6H-SiC(0001) substrates.  The controlled navigation and approach of ultra-small W-tips used as local ohmic contacts enables us to  directly correlate different  transport regimes with local structural properties of the graphene film. Furthermore, the detailed analysis of temperature dependency demonstrates, that the probe tips are able to provide well defined local contacts. Areas with qualitatively different graphene structures are hosted on the same SiC-sample, so that the results can be directly compared with each other.

\vspace{2ex}
Epitaxial graphene was grown on 6H-SiC(0001) surfaces by high temperature annealing in an Ar atmosphere \cite{Emtsev09,Starke12}. As intended for this work the growth procedure was stopped before the completion of large scale  graphene area in order to study different transport regimes by accessing patches of different graphene perfection. The characterization was done with a four-tip STM/SEM system (Omicron) operating at a base pressure of $\rm 2 \times 10^{-8}~Pa$ and temperatures down to 30~K. The high resolution SEM (4~nm, in-lens detector) allows the precise navigation of the four W-tips to desired positions in order to perform scanning tunneling microscopy (STM), scanning tunneling spectroscopy (STS) and local transport measurements.
In order to avoid irreversible damages to the probes and the graphene  all tips were approached first via individual  feedback controlled loops into a tunneling contact while ohmic contacts were realized by a defined push down of the tips with calibrated piezos in the  feedback-off mode.  STS data were taken using a lock-in technique ($\rm V_{ac}=20~meV, f=1.5~kHz$). The thickness and quality of the epitaxial graphene films were checked further by ex-situ confocal  Raman spectroscopy operating at 532~nm wavelength \cite{Graf07}. The overall electronic properties of graphene have been controlled by angle resolved photoemission  spectroscopy (ARPES) \cite{Starke12}.
\vspace{2ex}

\begin{figure}[tb]
\begin{center}
\includegraphics[width=0.65\columnwidth]{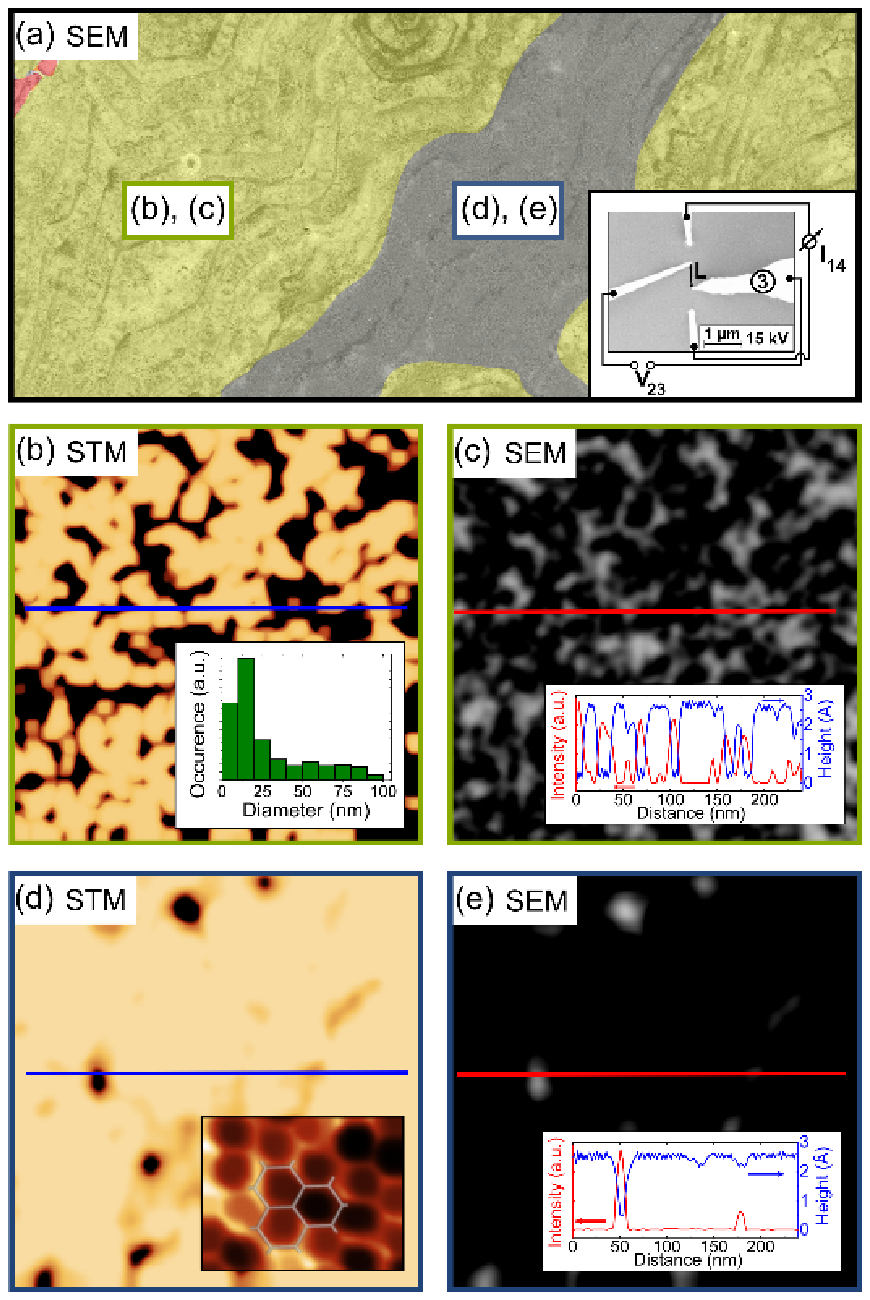}
\caption{\label{FIG1} (color online) a) Color-coded SEM large-scale image ($\rm 96 \times 45 \mu m^2$, 15~kV) of inhomogeneously grown graphene. The reddish area shows clean SiC(0001) areas.  The inset shows collinearly arranged  tips as used for transport measurements on the different graphene areas (cf. with Fig.~\ref{FIG3}). The STM and STS  data shown in  (b,d) and Fig.~\ref{FIG2} are taken with tip 3. b)-f): STM (b,d, +1.1~V, 0.5~nA) and SEM (c,e, 15~kV) images (scan area $\rm 250 \times 250~nm^2$) taken in the imperfect graphene (b,c greenish area in Fig.~\ref{FIG1}) and quasi-perfect part (d,e, greyish colored). The STM and SEM were taken in each area at the same location. The insets in c) and e) show line profiles taken in the according STM (blue) and SEM (red) images. Inset in b): histogram  of the high/low contrast of SEM images (in total $\rm 3 \times 3 ~\mu m^2$) taken in imperfect area. Inset in d): atomically resolved honeycomb lattice of graphene in the quasi-perfect area.}
\end{center} \end{figure}

Fig.~\ref{FIG1}a) shows a large-scale SEM image of an area with  inhomogeneously grown graphene. Besides very small patches of the bare substrate (reddish colored area), quasi-perfect graphene (greyish colored) as well as imperfect graphene (greenish area) are present whose properties are discussed in the following.

\begin{figure}[tb]
\begin{center}
\includegraphics[width=0.8\columnwidth]{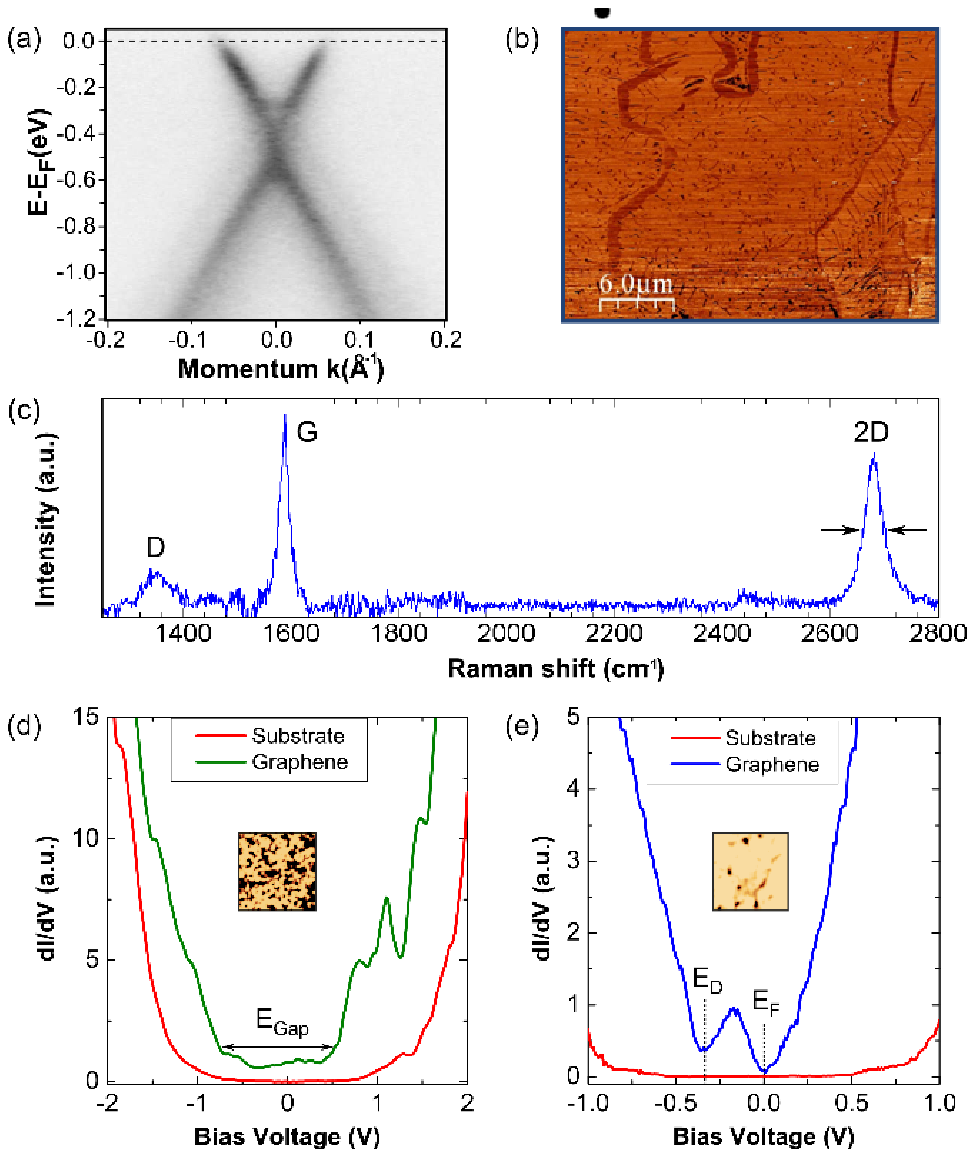}
\caption{\label{FIG2} (color online) ARPES (a) and AFM phase contrast image (b) of a quasi-perfect graphene area (cf. greyish area in Fig.~\ref{FIG1}) of a similarly prepared sample. The darker contrast at the step edges is due to bilayer fractions grown preferentially at step edges. c) Raman spectrum  (after substraction of the SiC-signal) of the sample shown in Fig.~\ref{FIG1}a). The position of the G- (1588 $\rm cm^{-1}$) and 2D-peaks (2680 $\rm cm^{-1}$) and their small widths at half maximum (21 $\rm cm^{-1}$ and 40 $\rm cm^{-1}$, respectively) confirm the growth of single layer graphene layers on SiC(0001). STS spectra taken on defected (d) and perfect graphene (e). The STM-images symbolize these areas.  For comparison in both graphs the spectrum of the SiC surface is shown for reference. The shift of the Dirac point in (e) reflects directly the n- doping level of about 360~meV in the perfect graphene area.}
\end{center} \end{figure}

Magnifications of the two characteristic graphene areas are shown in Fig.~\ref{FIG1}b)-e).
Thereby, the SEM and STM images  were taken at the same location in each of the two areas, thus allowing us to correlate directly the intensity variation in SEM with the height variation  as shown by the insets (panels c,e). The variation was confirmed by atomic force microscopy (AFM) thus excluding electronic effects which often mimic height variations in STM. The apparent heights of 2.3~\AA~ correlate nicely with the graphene step height seen for graphene on buffer-layer on SiC(0001) \cite{Rutter07} and shows that the high- and low-contrast seen in SEM is related to the buffer-layer and graphene phases, respectively. Furthermore, as supported by AFM (cf. with Fig.~\ref{FIG2}b), small fractions of bilayer are already formed at step edges appearing  also in the SEM image shown in Fig.~\ref{FIG1}a). However, the bilayer fraction is a minority and therefore does not influence the local transport properties.

The correlation of the SEM intensity with STM height information enables us to measure large-scale areas rapidly with a sufficient resolution for further statistical analysis.  In particular, the statistical analysis of the distance between high/low intensity ratio  (histogram in Fig.~\ref{FIG1}b) reveals an average graphene island diameter of 12~nm which is fully supported by temperature dependent transport measurements presented below. The corrugation of the (almost) perfect graphene area is around 0.2~\AA~ (inset in panel e) and matches with typical rms-roughnesses reported earlier \cite{Riedl07,Choi10}.

As we will show in the following  the high (low) intensity in SEM is directly linked to the insulating (conductive) behavior  of non-intact (intact) graphene. This strict anti-correlation of the SEM signal with the topography of graphene is very useful because it allows a fast and controlled navigation of the STM-tip to desired positions in order to perform spatially resolved STS and transport measurements which will be  discussed in detail in the following.

Spectroscopically, the samples have been characterized by ARPES, Raman and STS. An ARPES spectrum is shown in Fig.~\ref{FIG2}a) and  clearly reveals the signatures of graphene. The faint intensities around the Dirac cone stems from the bilayer fraction ($\rm < 10\%$) visible also in the AFM phase contrast image shown in Fig.~\ref{FIG2}b). In contrast to the STS-measurements the Raman signal was detected from arbitrary sites of the sample, i.e. integrating both perfect and imperfect areas of the sample.
A characteristic Raman spectrum of this surface (after substraction of the signal measured on a bare SiC substrate) is shown in Fig. 2c).  The G-line, centered at $\rm 1588~cm^{-1}$, as well as the 2D-line, centered at $\rm 2680~cm^{-1}$, with a full width at half maximum (FWHM) of $\rm 40~cm^{-1}$, are characteristic for monolayer graphene \cite{Emtsev09}. It should be noted that our values compare very well with those found for exfoliated graphene \cite{Ferrari13}.
However, the existence of a D-line at $\rm 1352~cm^{-1}$ indicates the presence of defects in the graphene \cite{Ferrari07}. In our samples this peak is most likely induced by the nano-inhomogeneities. Nonetheless, the Raman signal unambiguously proves the growth of monolayer graphene.

The structural imperfections are accompanied by electronic heterogeneities as easily probed by local STS.  The dI/dV signal taken at the bare substrate  (e.g. redish area in Fig.~\ref{FIG1}a) shows a band gap of almost 3~eV which is in reasonable agreement with the band gap of 6H-SiC(0001). The gap is significantly reduced on those areas where both STM and SEM have revealed the imperfect area (greenish area in Fig.~\ref{FIG1}). Extrapolation of the band edges reveal an effective  gap  around 1~eV which is close to the value reported for  graphene buffer layers on SiC(0001) \cite{Varchon07,Goler13}.  STS-measurements taken at different sites within this  imperfect area reveal similar spectra, i.e. a clear discrimination between locally intact and imperfect  graphene  has not been achieved. Instead, the STS spectra reveal a constant density of states within the transport gap as expected for defected graphene \cite{Peters12}. Different to the spectra on these patches, though, are spectra taken on quasi-perfect graphene (cf. with inset of Fig.~\ref{FIG2}e) revealing  the  characteristic signatures, i.e. the  linear band edges as well as the Dirac point which is shifted to negative voltages ($\rm E_D \approx -360~meV$) due to the intrinsic n-doping of graphene  grown on the Si-face of SiC(0001) surfaces in reasonable agreement with the ARPES shown in Fig.~\ref{FIG2}a) \cite{Ristein12}.

With these information as prerequisite the different nano-scaled graphene areas can be probed selectively regarding transport using the four-tip STM technique. A typical 4-tip collinear assembly used with various tip spacings L and substrate temperatures is shown in the inset of Fig.~\ref{FIG1}.
The sheet resistance $\rm R_{sheet}$ for a collinear and equidistant arrangement of the tips (the contact areas are around  $\rm 20 \times 20 ~nm^2$ and are small compared to L) can be calculated  via $\rm R_{sheet}=\pi / ln 2~V_{23}/I_{14}$, where $\rm I_{14}$  and $V_{23}$ are the current and measured voltage drop, respectively. The indexing denotes the tips (cf. with Fig.~\ref{FIG1}a, $\rm R_{23,14} \equiv V_{23}/I_{14}$).

\begin{figure}[tb]
\begin{center}
\includegraphics[width=1.03\columnwidth]{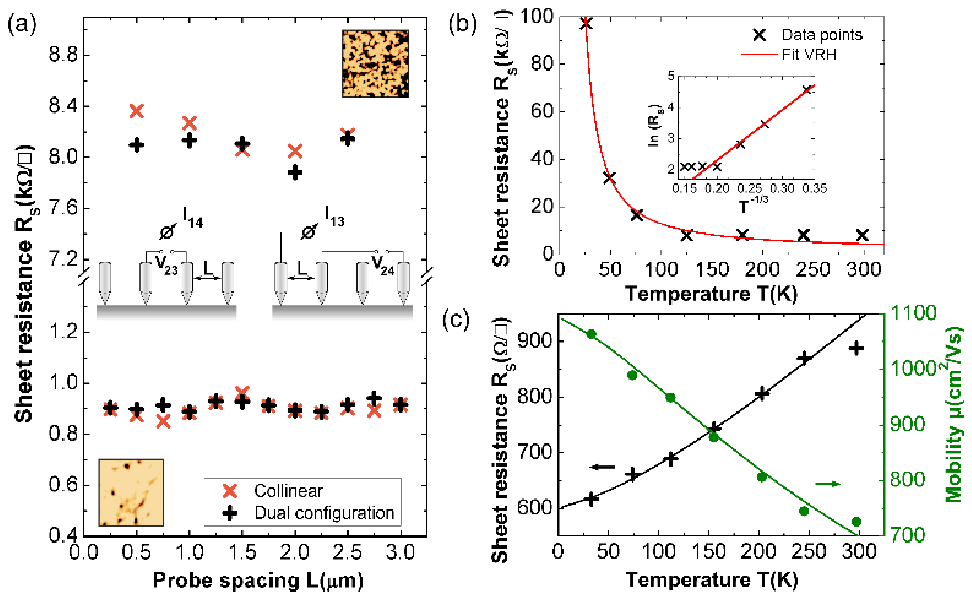}
\caption{\label{FIG3} (color online) Local transport measurements performed on imperfect and quasi-perfect graphene as function of probe spacings (a,T=300~K)  and temperature (b,c, probe spacing L=$\rm 1~ \mu m$). The inset in (a) shows the two configuration setups for transport measurements. The STM-images symbolize  the two different graphene areas where the transport measurements have been performed.}
\end{center} \end{figure}

The results obtained on both characteristic areas are shown in Fig.~\ref{FIG3}a) for a constant substrate temperature of 300~K. The sheet resistance  for the inhomogeneous areas is of the order of $\rm R_{sheet} \approx 8~k\Omega/\Box$ while it is lower by an order of magnitude for the quasi perfect area. The resistivity  is not varying with distance which  proofs two dimensional transport in both areas.

The calculation of the sheet resistance according to the formula given above is possible only if the tip spacing L is large compared to the film thickness and, at the same time, small compared to the lateral extension of the area which is probed.  Furthermore, the distance of the probes from the edges of the finite patches should be large compared to their spacings. Otherwise, edge effects  play a role and need to be considered by appropriate correction factors \cite{Schroder06}.
For this approach, the resistance has been measured in a further configuration - in addition to the above proposed tip configuration - namely as  $\rm R_{24,13}$. Thus the sheet resistance can be calculated via $\rm 1=exp(2 \pi R_{23,14}/R_{sheet})-exp(2 \pi R_{24,13}/R_{sheet})$ without relying on correction factors \cite{Wang10}.  The deviation from the ideal value, i.e.  $\rm \pi/ln2$, was found to be $\rm < 1$~\% (2~\%) for the imperfect (quasi-perfect) graphene area. The quantitative agreement of both data sets  demonstrates  that due to the spatial control of the tip placement  finite size effects  are not dominant and, finally, enables us to probe two adjacent areas with different transport regimes on the same sample.

In order to further quantify the transport regimes temperature dependent measurements were performed. For the imperfect area when decreasing temperature the resistance increases exponentially up to $\rm 100~k \Omega/\Box$ at $\rm T=30~K$ (cf. Fig.~\ref{FIG3}b) strongly indicating activated transport. The probe spacing was kept fixed at $\rm L=1~\mu m$.

This change of  resistivity with temperature  up to 200~K is reliably described by Anderson localization, i.e. $\rm R_{sheet} \propto exp((T_0(\xi)/T)^{1/d+1})$ where $\rm d=2$ denotes the dimension (cf. with inset in Fig.~\ref{FIG3}b) \cite{Mott71}. The deviation above 200~K might be attributed to phonon-assisted contributions which are not considered further here in this transport regime. According to Mott $\rm T_0(\xi) \approx 14(\xi^2 k_B D(E_F))^{-1}$ where $\rm \xi$ denotes the localization length, $k_B$ the Boltzmann factor and $\rm D(E_F)$ the density of (defect) states at the Fermi energy\cite{Tsui74,Skal74}.  The fit to our experimental data reveal an Anderson temperature of $\rm T_0=4200~K$, which corresponds to a localization length of around $\rm \xi \approx 12~nm$ (see histogram in Fig.~\ref{FIG1}b) if  $\rm D(E_F) \approx  2 \times 10^{13} cm^{-2} (eV)^{-1} $ is assumed.  The transport mechanism in this defected graphene and description in terms of variable range hopping (VRH)  supports  the conclusions drawn from recent  large-scale transport  experiments performed on graphene antidot lattice structures \cite{Zhang13, Peters12}.

Quasi-perfect graphene reveals a different behavior.
The temperature dependence of the resistance in the alleged perfect graphene area was measured and the results are  shown in Fig.~\ref{FIG3}c). In contrast to the imperfect area, the sheet resistance increases  with increasing temperature which clearly rules out transport via VRH. In fact the temperature dependence of the resistivity can be accurately described  by considering  phonon assisted scattering in addition to a  residual background $\rm \rho_0$. In recent studies it has been demonstrated  that, besides contributions from longitudinal acoustic phonons in graphene ($\rm \rho_{LA}$), also activated contributions from low-energy  phonon modes ($\rm E_1=70~meV$ and $\rm E_2=16~meV$) need to be considered ($\rm \rho_P=\sum_{i=1}^2 C_i [exp(E_i/k_B T)-1]^{-1}$) \cite{Giesbers12, Chen08}. As deduced from inelastic tunneling spectroscopy performed on graphite \cite{Vitali04}, these modes can be related to acoustic out-of plane modes. The analysis demonstrates that the graphene layer is indeed coupled to the underlying buffer-layer which is closely related to the scenario of remote interfacial phonon  scattering \cite{Chen08}. Finally, as we know precisely the chemical potential from STS  the electron mobility $\rm \mu$ can be calculated from the resistivity. The mobilities in our case  were found to be around 700 $\rm cm^2/Vs$ at room temperature in reasonable agreement with measured mobilities \cite{Speck11,Jobst11}.

In general our results compare well with those reported from other groups, however,  the temperature dependence  is weaker by a factor of 2  \cite{Speck11, Giesbers12,Jobst11}. For instance, the electron-phonon coupling constants for the two phonon modes $\rm E_1$ and $\rm E_2$ in the $\rm \rho_P$ contribution were found to be $\rm C_1=282~\Omega$ and $\rm C_2=112~\Omega$, respectively, which is significantly lower than those values reported in Ref.~\cite{Giesbers12}.
The contribution of $\rm \rho_{LA}$ (for details see e.g. Ref.~\cite{Giesbers12}) is closely related to the so-called deformation potential $\rm D_A$, which accounts for strain or phonon-induced  changes on  the electronic band structure. The resistivity curve shown in Fig.~\ref{FIG3}c)  is described best with $\rm D_A=18~eV$ which is close to the theoretically expected value \cite{Suzuura02} and which was found for instance in exfoliated graphene samples on $\rm SiO_2$ \cite{Chen08}. In contrast for graphene on SiC larger values for $\rm D_A$ are reported which are thought to be induced by additional strain effects \cite{Giesbers12}. Therefore, we conclude that the on-top nano-sized contacts used in our experiments do  not induce significant stress  that would result in larger coupling constants and deformation potentials.

In summary, we have presented a comprehensive study of micron-sized graphene grown on SiC(0001). Besides the morphology the local density of states has been measured and correlated with local transport measurements. The analysis of transport as a function of probe spacings and probe geometries has shown that the graphene patches can be probed locally and that the current paths are not influenced by finite size effects of the different areas. From temperature dependent measurements the regimes of VRH and diffusive transport have been identified. In particular, the detailed analysis of  the latter transport regime has shown that the contacts do  not disturb the graphene system.  Furthermore the SEM/STM capability of our instrument has been  used to calibrate the morphology seen in STM with intensity variations in SEM, thus enabling us to position  the probes reliably to desired positions on the sample. Our results show that both positioning of the contacts as well as a gentle contacting are decisive for transport measurements on low dimensional structures.


\vspace{1ex}

{\bf Acknowledgement} Financial support by the Deutsche
Forschungsgemeinschaft in the framework of the Priority Program
SPP 1459 "Graphene" is gratefully acknowledged.\\



\begin{thebibliography}{99}
\bibitem{Novoselov12}
K.S. Novoselov, F.I. Fal'ko, L. Clombo, P.R. Gellert, M.G. Schwab, and K. Kim, Nature {\bf 490}, 192 (2012).
\bibitem{Emtsev09}
K. V. Emtsev, et.al.,
Nat. Mater. {\bf 8}, 203 (2009).
\bibitem{Reina09}
A. Reina, X. Jia, J. Ho, D. Nezich, H. Son, V. Bulovic, M. S. Dresselhaus, and J. Kong, Nano Lett. {\bf 9}, 30 (2009).
\bibitem{Berdebes11}
D. Berdebes, T. Low, Y. Sui, J. Appenzeller,  and M.S. Lundstrom, IEEE Trans. Elec. Dev. {\bf 58}, 3925 (2011).
\bibitem{Martin08}
J. Martin, N. Akerman, G. Ulbricht, T. Lohmann, J. H. Smet, K. von Klitzing, and A. Yacoby, Nature Physics {\bf 4}, 144 (2008).
\bibitem{Miller09}
D. L. Miller, K. D. Kubista, G. M. Rutter, M. Ruan, W. A. de Heer, P. N. First, and J. A. Stroscio, Science {\bf 324}, 924 (2009).
\bibitem{Weiss10}
C. Weiss, C. Wagner, C. Kleimann, M. Rohlfing, F.S. Tautz, and R. Temirov, Phys. Rev. Lett. {\bf 105}, 086103 (2010).
\bibitem{Gross11}
L. Gross, Nat. Chem. {\bf 3}, 273 (2011).
\bibitem{Welker12}
J. Welker and  F. J. Giessibl, Science {\bf 336}, 444 (2012).
\bibitem{Starke12}
U. Starke, S. Forti, K.V. Emtsev, and C. Coletti, MRS Bulletin {\bf 37}, 1177 (2012).
\bibitem{Graf07}
D. Graf, F. Molitor, K. Ensslin, C. Stampfer, A. Jungen, C. Hierold, and L. Wirtz, Nano Lett. {\bf 7}, 238 (2007).
\bibitem{Rutter07}
G. M. Rutter, N. P. Guisinger, J. N. Crain, E. A. A. Jarvis, M. D. Stiles, T. Li, P. N. First, and J. A. Stroscio, Phys. Rev. B {\bf 76}, 235416 (2007).
\bibitem{Riedl07}
C. Riedl, U. Starke, J. Bernhardt, M. Franke, and K. Heinz, Phys. Rev. B {\bf 76}, 245406 (2007).
\bibitem{Choi10}
J. Choi, H. Lee, and S. Kim, J. Phys. Chem. C {\bf 114}, 13344 (2010).
\bibitem{Ferrari13}
A.C. Ferrari and D.M. Basko, Nat. Nanotechnology {\bf 8}, 235 (2013).
\bibitem{Ferrari07}
A.C. Ferrari, Solid Stat. Comm. {\bf 143}, 47 (2007).
\bibitem{Varchon07}
F. Varchon et.al.,
 Phys. Rev. Lett. {\bf 99}, 126805 (2007).
\bibitem{Goler13}
S. Goler et. al. Carbon {\bf 51}, 249 (2013).
\bibitem{Peters12}
E.C. Peters, A.J.M. Giesbers, and M. Burghard Phys. Status Solidi B {\bf 249}, 2522 (2012).
\bibitem{Ristein12}
J. Ristein, S. Mammadov, and T. Seyller, Phys. Rev. Lett. {\bf 108}, 246104 (2012).
\bibitem{Schroder06}
D. K. Schroder, Semiconductor material and device characterization, John Wiley und Sons, New Jersey, 2006.
\bibitem{Wang10}
Wang et al., J. Vac. Sci. Tech. B {\bf 28}, C1C34 (2010).
\bibitem{Mott71}
N. F. Mott and E. A. Davis, Electronic Processes in Noncrystalline Materials, Oxford Univ. Press, London, 1971.
\bibitem{Tsui74}
D.C. Tsui and S.J. Allen, Jr., Phys. Rev. Lett. {\bf 32}, 1200 (1974).
\bibitem{Skal74}
A.S. Skul and B.I. Shklovskii. Sov. Phys. Solid State {\bf 16}, 1190 (1974).
\bibitem{Zhang13}
H. Zhang et. al., Phys. Rev. Lett. {\bf 110}, 066805 (2013).
\bibitem{Giesbers12}
A.J.M. Giesbers, P. Proch$\rm \acute{a}$zka, C.F.J. Flipse, Phys. Rev. B {\bf 87}, 195405 (2012).
\bibitem{Chen08}
J.-H. Chen, C. Jang, S. Xiao, M. Ishigami, and M. S. Fuhrer,
Nat. Nano. {\bf 3}, 206 (2008).
\bibitem{Vitali04}
L. Vitali, M. A. Schneider, K. Kern, L. Wirtz, and A. Rubio, Phys. Rev. B {\bf 69}, 121414(R) (2004).
\bibitem{Jobst11}
J. Jobst, D. Waldmann, F. speck, R. Hirner, D.K. Maude, Th. Seyller, and H.B. Weber, Sol. State. Comm. {\bf 151}, 1061 (2011).
\bibitem{Speck11}
F. Speck, J. Jobst, F. Fromm, M. Ostler, D. Waldmann, M. Hundhausen, H. B. Weber, and Th. Seyller, Appl. Phys. Lett. {\bf 99}, 122106 (2011).
\bibitem{Suzuura02}
H. Suzuura and T. Ando, Phys. Rev. B {\bf 65}, 235412 (2002).
\end{thebibliography}
\end{document}